\documentclass[aps,pre,nofootinbib,showpacs,twocolumn,10pt]{revtex4-1} 

\usepackage{amssymb,latexsym,mathrsfs}
\usepackage{amsfonts}
\usepackage{amsmath}
\usepackage{graphicx}
\usepackage{srcltx}
\usepackage{color}
\usepackage{cancel}

\begin{document}

\title{Phase transition in an exactly solvable reaction-diffusion process}
\author{Somayeh Zeraati$^1$, Farhad H. Jafarpour$^1$, and Haye Hinrichsen$^2$}
\affiliation{$^1$Bu-Ali Sina University, Physics Department, 65174-4161 Hamedan, Iran}
\affiliation{$^2$Universit\"at W\"urzburg, Fakult\"at f\"ur  Physik und Astronomie, 97074  W\"urzburg, Germany,}
\email{email: s.zeraati@basu.ac.ir\\farhad@ipm.ir\\hinrichsen@physik.uni-wuerzburg.de}

\def\d{{\rm d}}
\def\0{\emptyset}
\def\pi{p^{(i)}}
\def\pf{p^{(\infty)}}

\def\comment#1{\color{red}[ Comment: #1]\color{black}}
\def\mark#1{\color{red}#1 \color{black}}

\begin{abstract}
We study a non-conserved one-dimensional stochastic process which involves two species of particles $A$ and~$B$. The particles diffuse asymmetrically and react in pairs as $A\emptyset\leftrightarrow AA\leftrightarrow BA \leftrightarrow A\emptyset$ and $B\emptyset \leftrightarrow BB \leftrightarrow AB \leftrightarrow B\emptyset$. We show that the stationary state of the model can be calculated exactly by using matrix product techniques. The model exhibits a phase transition at a particular point in the phase diagram which can be related to a condensation transition in a particular zero-range process. We determine the corresponding critical exponents and provide a heuristic explanation for the unusually strong corrections to scaling seen in the vicinity of the critical point.
\end{abstract}

\maketitle
\parskip 1mm


\section{Introduction}

One-dimensional driven-diffusion systems have been a subject of study in recent years because they exhibit interesting properties such as non-equilibrium phase transitions~\cite{Schmittmann}. These systems have many applications in different fields of physics and biology~\cite{Schutz,MacdonaldGibbsPipkin}. A well-known example is the Asymmetric Simple Exclusion Process (ASEP), which is studied experimentally by optical tweezers~\cite{Optical tweezers1,Optical tweezers2}. 

Various approaches have been developed in order to solve such systems exactly, including for example the matrix product method. With the matrix product method, the steady-state weight of a configuration is written as the trace of a product of operators corresponding to the local state of each lattice site. The operators obey certain algebraic rules which are derived from the dynamics of the model~\cite{DEHP}. The algebraic relations among these operators might have finite or infinite dimensional matrix representations~\cite{EsslerRittenberg,BlytheMPM}. Recently the matrix product method with quadratic algebras attracted renewed attention as it can also be applied to dissipative quantum systems~\cite{Prosen,Karevski}.

It is well known that one-dimensional systems with open boundary conditions, in which the particle number is not conserved at the boundaries, can exhibit a phase transition~\cite{EvansNonconserving}. On the other hand a phase transition may also take place in systems with non-conserving dynamics in the bulk~\cite{Hinrichsen,EvansKafri}. For example, in Ref.~\cite{EvansKafri} the authors have studied a three-states model on a lattice with periodic boundary conditions with two particle species which evolve by diffusion, creation and annihilation. By changing the annihilation rate of the particles, this model displays a transition from a maximal current phase to a fluid phase.

As shown in~\cite{EvansZRP1} it is possible to map a one-dimensional driven-diffusive system defined on a periodic lattice onto a so-called zero-range process (ZRP). Recently this mapping was used to study various models which have an exact solution in the steady state~\cite{EvansZRP,KafriZRP}. It was shown that a phase transition in the original model corresponds to a condensation transition in the corresponding ZRP.

In present work, we introduce and study an exactly solvable one-dimensional driven-diffusive model with non-conserved dynamics which exhibits an interesting type of phase transition. The model is defined  on a ring of $L$ sites which can be either empty (denoted by a vacancy $\emptyset$) or occupied by a one particle of type $A$ or type $B$. The system evolves random-sequentially according to a set of two-site processes which can be written in the most general form as
\begin{equation}
\label{DYNAMICAL RULE}
I\emptyset\mathop{\rightarrow}\limits^{\alpha_{I}} \emptyset I\,,\qquad
IK\mathop{\rightleftharpoons}\limits^{\beta_{IJ}}_{\beta_{JI}} JK\,,\qquad
IJ\mathop{\rightleftharpoons}\limits^{\omega_{IJ}}_{\omega_{JI}} J\emptyset\,,
\end{equation}
where $I, J, K\in \{A,B\} $. In what follows we study a special case of this model defined by the processes
\begin{equation}
\label{DYNAMICALRULES}
\begin{array}{cc}
A\emptyset\mathop{\rightarrow}\limits^{\alpha_{+}} \emptyset A,&
B\emptyset\mathop{\rightarrow}\limits^{\alpha_{-}}  \emptyset B\\
AB\mathop{\rightleftharpoons}\limits^{p_{+}}_{p_{-}} BB,&
AA\mathop{\rightleftharpoons}\limits^{p_{+}}_{p_{-}} BA\\
AB\mathop{\rightleftharpoons}\limits^{\alpha_{+}}_{\alpha_{-}} B\emptyset,&A\emptyset\mathop{\rightleftharpoons}\limits^{p_{+}}_{p_{-}} BA\\
A\emptyset\mathop{\rightleftharpoons}\limits^{1}_{1} AA,&
B\emptyset\mathop{\rightleftharpoons}\limits^{p}_{\alpha} BB\\
\end{array}
\end{equation}
where the rates $\alpha$ and $p$ are given by the ratios
\begin{equation}
\label{ratios}
\alpha=\frac{\alpha_+}{\alpha_-}\,,\qquad p = \frac{p_+}{p_-}\,.
\end{equation}
As we will see below, for this particular choice the model turns out to be exactly solvable. Obviously, this defines a non-conserved dynamics, allowing the number of particles ($N_{A}$ and $N_{B}$) and vacancies ($N_{\emptyset}$) to fluctuate under the constraint $L=N_{A}+N_{B}+N_{\emptyset}$. Moreover, the model is a driven system since diffusion and reaction processes are not left-right symmetric. The dynamical rules~(\ref{DYNAMICAL RULE}) is extensible to an exactly solvable model with the various types of particles, in which a phase transition is accessible. A  generalized model consisting of three species of particles is presented in the Appendix A.

In this paper we demonstrate that the model defined in~(\ref{DYNAMICALRULES}) exhibits a phase transition and that its stationary state can be determined exactly by means of the matrix product method. In Sect.~\ref{sec:zrp} we show that our model can be mapped onto a ZRP and that the phase transition corresponds to a condensation transition in the ZRP. In Sect.~\ref{sec:numerics} we study the dynamical behavior, which is not part of the exact solution, by numerical simulations. It turns out that the dynamical behavior near the critical point is plagued by unusually persistent corrections to scaling, which are explained from a phenomenological point of view in Sect.~\ref{sec:toymodel}.


\section{Phase diagram and phenomenological properties}

\begin{figure}
\centering\includegraphics[width=\linewidth]{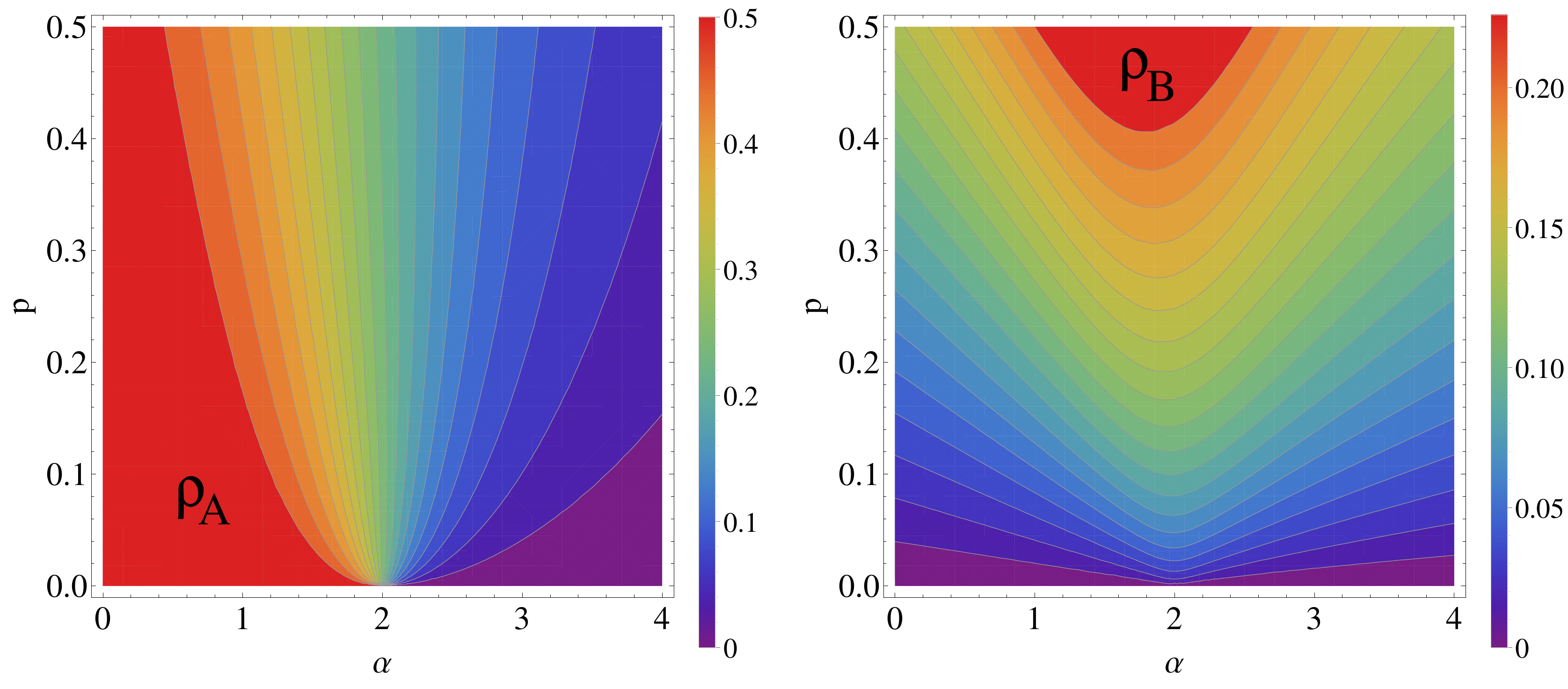}
\vspace*{-5mm}
\caption{Stationary density of the reaction-diffusion process investigated in the present work. Left: At the lower boundary $p=0$ the model exhibits two different phases, namely a high-density phase for $\alpha<2$ (red) and a low-density phase for $\alpha > 2$ (violet), separated by a discontinuous phase transition when moving along the bottom line in the left panel. For $p>0$ the order parameter $\rho_A$ changes continuously without exhibiting a phase transition. Right: The order parameter $\rho_B$ displays instead a continuous phase transition. }
\label{fig:rho}
\end{figure}

\label{sec:Phase diagram}
The model defined above is controlled by four parameters $\alpha_+$, $\alpha_-$, $p_+$, and $p_-$. As we will see below, the essential quantities which determine the matrix algebra are the ratios $\alpha=\frac{\alpha_+}{\alpha_-}$ and $p = \frac{p_+}{p_-}$ in Eq.~(\ref{ratios}), and therefore it is useful to study the phase diagram of the model in terms of these ratios. For the remaining two degrees of freedom we choose $\alpha_+\alpha_-=p_-=1$ throughout this paper, i.e. we use the definition
\begin{equation}
\alpha_+=\sqrt{\alpha}\,,\quad \alpha_-=\frac{1}{\sqrt{\alpha}}\,,\qquad p_+=p\,,\quad p_-=1.
\end{equation}

This selects a 2D subspace in the 4D parameter space which is believed to capture the essential phase structure of the system. 

The phase diagram for the particle densities $\rho_A$ and $\rho_B$ in terms of $\alpha$ and $p$ is shown in Fig.~\ref{fig:rho}. As can be seen, these densities vary continuously everywhere except for the point $\alpha=2, p=0$, where the model exhibits a phase transition. Moving along the horizontal axis at $p=0$, the order parameter $\rho_A$ jumps discontinuously from $1/2$ to $0$, indicating first-order behavior, while $\rho_B$ changes continuously as in a second-order phase transition.

\begin{figure*}
\centering\includegraphics[width=0.9\linewidth]{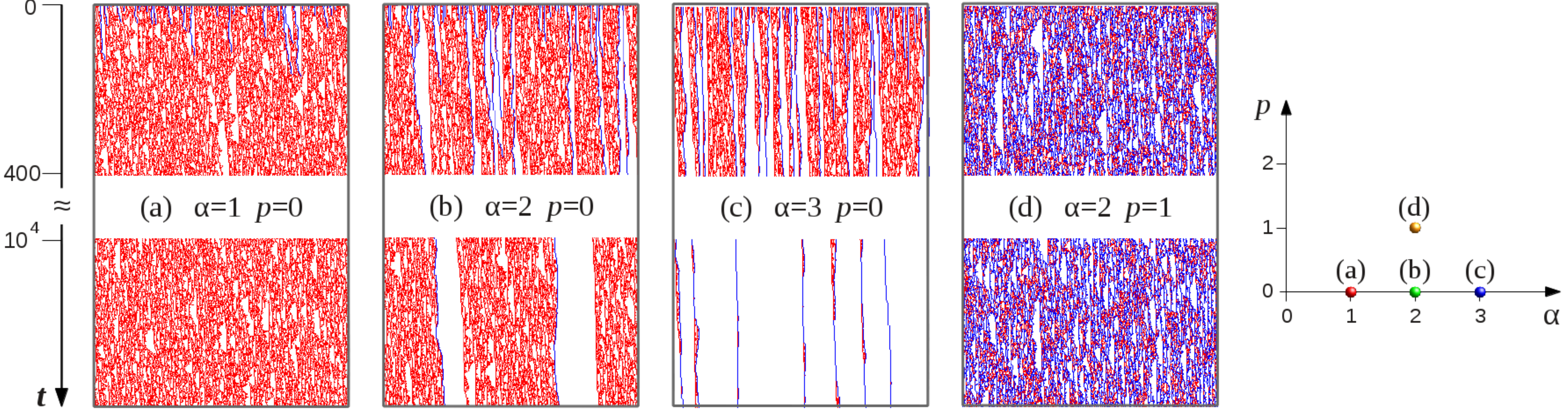}
\caption{Snapshots of typical space-time evolutions starting with random initial conditions. Particles of type $B$ are represented by blue pixels while $A$-particles are plotted in red color. The figure shows snapshots for four different choices of the parameters, corresponding to the points in the phase diagram shown on the right.}
\label{fig:demo}
\end{figure*}

To give a first impression how the process behaves in different parts of the phase diagram, we show various typical snapshots of the space-time evolution in Fig.~\ref{fig:demo}. For $p
=0$ the density of $B$-particles (blue pixels) is very low while the $A$-particles (red pixels) form fluctuating domains with a high density. As we will see in the last section, these sharply bounded domains are important for a qualitative understanding of the phase transition.

For $\alpha<2$ the $A$-particles eventually fill  the entire system while for $\alpha>2$ the $A$-domains almost disappear, leaving diffusing $B$-particles behind. For $p>0$ one can see that $B$-particles are continuously generated. Thus the parameter $\alpha$ controls the domain size of $A$-particles while the parameter $p$ controls the creation and therewith the density of $B$-particles.

\section{Exact results}

\label{sec:exact}
The matrix product method is an important analytical tool developed in the 90's to compute the steady-state of driven diffusive systems exactly~\cite{DEHP,BlytheMPM}. Let us now investigate the stationary state of the model by using this method. We consider a configuration $C=\{\tau_{1},\cdots,\tau_{L}\}$ with $\tau_{i}\in \{ \emptyset,A,B\}$ on a discrete lattice of length $L$ with periodic boundary condition. According to this method, the stationary state weight of a configuration $C$ is given by the trace of a product of non-commuting operators~$X_{i}$:
\begin{equation}
\label{Weights MPM}
W(C)={\rm Tr}\bigl[\prod_{i
=1}^{L}{X_{i}}\bigr]\,.
\end{equation}
Note that this method differs from the well-known transfer matrix method in so far as different matrices are used depending on the actual configuration of the lattice sites, i.e. the choice of the operator $X_{i}$ at site $i$ depends on its local state. In our model, the operator $X_{i}=\mathbf E$ stands for a vacancy while $X_{i}=\mathbf A  (\mathbf B)$ represents a particle of type $A$ ($B$). Depending on the dynamical rules, these operators should satisfy a certain set of algebraic relations. For the dynamical rules listed in~(\ref{DYNAMICALRULES}) one obtains a quadratic algebra of the form
\begin{equation}
\label{ALGEBRA}
\begin{array}{l}
p_{-}\mathbf B \mathbf A+\mathbf A \mathbf E=(1+p_{+})\mathbf A\mathbf A\\
p_{-}\mathbf B \mathbf B+\alpha_{-}\mathbf B \mathbf E=(\alpha_{+}+p_{+})\mathbf A\mathbf B\\
p_{+}\mathbf A \mathbf A+p_{+}\mathbf A \mathbf E=2p_{-}\mathbf B \mathbf A\\
p_{+}\mathbf A\mathbf B+p\mathbf B\mathbf E=(p_{-}+\alpha)\mathbf B\mathbf B\\
p_{-}\mathbf B\mathbf A+\mathbf A\mathbf A-(\alpha_{+}+p_{+})\mathbf A\mathbf E=\mathbf A\overline{\mathbf E}\\
\alpha_{+}\mathbf A\mathbf B+\alpha \mathbf B\mathbf B-(p+2\alpha_{-}-1)\mathbf B\mathbf E=\mathbf B\overline{\mathbf E}\\
\alpha_{+}\mathbf A\mathbf E-\mathbf E\mathbf A=- \overline{\mathbf E} \mathbf A\\
\alpha_{-}\mathbf B\mathbf E-\mathbf E\mathbf B=-\overline{\mathbf E}\mathbf B\\
\overline{\mathbf E}\mathbf E-\mathbf E\overline{\mathbf E}=0\end{array}
\end{equation}
where $\overline{\mathbf E}$ is an auxiliary matrix which is expected to cancel out in the final result. We find that the algebra~(\ref{ALGEBRA}) has a two-dimensional matrix representation given by the following matrices 
\begin{equation}
\label{MATRIX A,B,E}
\mathbf A=\left(
\begin{array}{cc}
1 & 0 \\
1 & 0
\end{array} \right),\;
\mathbf B=p\left(
\begin{array}{cc}
0 & 1 \\
0 & 1
\end{array} \right),\;
\mathbf E=\left(
\begin{array}{cc}
1 & 0 \\
0 & \alpha
\end{array} \right)
\end{equation}
and $\overline{\mathbf E}={\mathbf E}-\alpha_{+}{\mathbf I}$, where ${\mathbf I}$ is an identity $2\times2$ matrix. 

We note that the algebra (\ref{ALGEBRA}) and its representation (\ref{MATRIX A,B,E}) were studied for the first time by Basu  and Mohanty in Ref.~\cite{Basu} in the context of a different model. It differs from our one in so far as it evolves only according to the processes in the first two lines of (\ref{DYNAMICALRULES}), where the $A$ and $B$-particles hop with different rates and can also transform into each other, meaning that the total number of particles is conserved. The authors calculated the spatial correlations exactly and mapped their model to a ZRP. However, as the particle number is conserved in their model, a phase transition does not occur by changing the rates. In other words, although the matrix algebra already contains information about the phase transition, their model could not access the part of the phase diagram where the transition takes place. The model presented here is an extension of their model with the same matrix representation but with a non-conserved dynamics and an extended parameter space, in which the phase transition becomes accessible.

To compute the partition sum of the system, we first note that according to (\ref{DYNAMICALRULES}) a configuration without a particle of type $A$ or $B$ is not dynamically accessible. Therefore, the partition function, defined as a sum of the weights of all available configurations with at least one particle, is given by
\begin{equation}
\label{PartitionFunction}
Z_{L}={\rm Tr}\bigl[(\mathbf A+\mathbf B+\mathbf E)^L-\mathbf E^L\bigr]\,.
\end{equation}

With this partition sum the stationary density of the  $A$~and $B$-particles can be written as
\begin{equation}
\label{A particlesDensity}
\rho_A^{stat}=\frac{{\rm Tr}\bigl[\mathbf A(\mathbf A+\mathbf B+\mathbf E)^{L-1}\bigr]}{Z_{L}},
\end{equation}
\begin{equation}
\label{B particlesDensity}
\rho_B^{stat}=\frac{{\rm Tr}\bigl[\mathbf B(\mathbf A+\mathbf B+\mathbf E)^{L-1}\bigr]}{Z_{L}}\,.
\end{equation}

We can also compute the density of the vacancies using $\rho_{\emptyset}^{stat}=1-(\rho_A^{stat}+\rho_B^{stat})$. Using the representation~(\ref{MATRIX A,B,E}) the equations~(\ref{PartitionFunction})-(\ref{B particlesDensity}) can be calculated exactly. In the thermodynamic limit $L\mathop{\rightarrow}\infty$, where high powers of matrices are dominated by their largest eigenvalue, the density of the $A$ and $B$-particles is given by (see Fig.~\ref{fig:rho})
 \begin{equation}
\label{A particlesDensityTrmo}
\rho_A^{stat}= \frac{(2-\alpha)(\alpha+p)+\alpha \sqrt{4-4\alpha+(p+\alpha)^2} }{2(2\alpha+p)\sqrt{4-4\alpha+(p+\alpha)^2}},
\end{equation}
\begin{equation}
\label{B particlesDensityTrmo}
\rho_B^{stat}=\frac{p(p+3\alpha-2+\sqrt{4-4\alpha+(p+\alpha)^2})}{2(2\alpha+p)\sqrt{4-4\alpha+(p+\alpha)^2}}\,.
\end{equation}
Approaching the critical point at $p=0$ and $\alpha_{c}=2$, we find a discontinuous behavior 
\begin{equation}
\label{A particlesDensityP=0}
\rho_A^{stat}=
\left\{
\begin{array}{ll}
\frac{1}{2}   
&  \mbox{for} \; \alpha <\alpha_{c}\\ & \\
0
& \mbox{for} \;  \alpha >\alpha_{c}
\end{array}
\right.
\end{equation}
\begin{equation}
\label{VacanciesDensityP=0}
\rho_{\emptyset}^{stat}=
\left\{
\begin{array}{ll}
\frac{1}{2}   
&  \mbox{for} \; \alpha <\alpha_{c}\\ & \\
1
& \mbox{for} \;  \alpha >\alpha_{c}
\end{array}
\right.
\end{equation}

while $\rho_B^{stat}=0$. In fact, it is clear from  ($\ref{DYNAMICALRULES}$) that for $p=0$, the $B$-particles can only transform into $A$-particles or vacancies but they are not created. Hence, in the steady state in the thermodynamic limit, the $B$-particles will disappear.

We also observe that the density of the $B$-particles in the vicinity of the critical point changes discontinuously in a particular limit. This can be seen already in the snapshots of Fig.~\ref{fig:demo}a and ~\ref{fig:demo}c: For $\alpha<2$ and $p=0$ the density of $B$-particles vanishes rapidly on an exponentially short time scale, while for $\alpha>2$ one observes some kind of annihilating random walk with a slow algebraic decay. Therefore, for a small value of $p>0$, i.e. when switching on the creation of $B$-particles at a small rate, it is plausible that the system will respond differently in both cases. In fact, expanding (\ref{B particlesDensity}) around $p=0$ to first order in $p$  in the two phases $\alpha>\alpha_{c}$ or $\alpha
=\alpha_{c}+\epsilon $ and $\alpha<\alpha_{c}$ or $\alpha
=\alpha
_{c}-\epsilon $, where $\epsilon $ is very small, we find a band gap as
\begin{equation}
\label{BandGap}
\Delta =\rho_B^{stat ,\alpha>\alpha_{c}}-\rho_B^{stat ,\alpha<\alpha_{c}}\approx \frac{L^{2}p \epsilon}{8}
\end{equation}
which is valid for  $1 \ll L\ll L_{max}$ where $L_{max}=(p\epsilon)^{-1/2}$.


\section{Relation to a zero-range process}

\label{sec:zrp}
A  zero range process (ZRP) is defined as a system of $L$ boxes where each box can be empty or occupied by an arbitrary number of particles. The particles hop between neighboring boxes with a rate that can depend on the number of particles in the box of departure \cite{EvansZRP}. The stationary state of the ZRP factorizes, meaning that the steady-state weight of any configuration is given by a product of factors associated with each of the boxes. 

It is known that various driven-diffusive systems can be mapped onto a ZRP \cite{EvansZRP}. This is usually done by interpreting the vacancies (particles) in the driven-diffusive systems as particles (boxes) in the ZRP. Following the same line we find that our model can be mapped onto a non-conserving ZRP with two different types of boxes. More specifically, the $n$ vacancies to the right of an $A$($B$)-particle are regarded as an $A$($B$)-box containing $n$ particles in the ZRP denoted as $A_n(B_n)$. The total number of particles distributed among the boxes is denoted as~$N_\emptyset$ while number of boxes of type $A$($B$) is denoted as $N_A$($N_B$). By definition, the sum $N_{A}+N_{B}+N_{\emptyset}=L$ is conserved. However, the individual numbers are not conserved and change according to the following dynamical rules:
\begin{enumerate}
\item[(i)]
Particles from an $A$($B$)-box hop to the neighboring left box with rate $\alpha_{+}$ ($\alpha_{-}$):
  \begin{align}
    \label{RulesZRP}
  X_mA_n  &\mathop{\longrightarrow}\limits^{\alpha_+} X_{m+1}A_{n-1}\\ 
  X_mB_n  &\mathop{\longrightarrow}\limits^{\alpha_-} X_{m+1}B_{n-1} \qquad (X=A,B) \notag
  \end{align}
\item[(ii)]
An empty $A$($B$)-box transforms into an empty $B(A)$ box with the rate  $p
_{+}$ ($p_{-}$):
  \begin{equation}
  A_0  \mathop{\rightleftharpoons}\limits^{p_+}_{p_-} B_0   
  \end{equation}
\item[(iii)]
An $A$($B$)-box with $n$ particles together with an adjacent empty $B$($A$)-box on the left side transforms into a single $A$($B$)-box containing $n+1$  particles with rate  $p_{-}$ ($\alpha_{+}$). The reversed process is also possible and takes place with rate $p_{+}$ ($\alpha_{-}$):
  \begin{equation}
   B_0A_n \mathop{\rightleftharpoons}\limits^{p_-}_{p_+} A_{n+1}\,,\qquad
   A_0B_n \mathop{\rightleftharpoons}\limits^{\alpha_+}_{\alpha_-} B_{n+1} 
   \end{equation}
\item[(iv)]
 An $A$($B$)-box containing $n$ particles and a neighboring empty $A(B)$-box on the left side transform into an $A$($B$)-box with $n+1$  particles with the rate $1$ ($\alpha$). The reversed process is also possible and takes place with rate $1$ ($p$):
   \begin{equation}
   \label{EndRulesZRP}
   A_0 A_n \mathop{\rightleftharpoons}\limits^{1}_{1} A_{n+1}\,,\qquad
   B_0 B_n \mathop{\rightleftharpoons}\limits^{\alpha}_{p} B_{n+1}
   \end{equation}

\end{enumerate}
With these dynamical rules, we can show that the weights of configurations in the ZRP can be expressed as  factorized forms. We consider a configuration consisting of $\delta=N_{A}+N_{B}$ boxes with $N_{\emptyset}$ particles distributed in the boxes. Defining $n_k$ as the number of particles in $k^{\rm th}$ box of type $\tau_{k}\in \{A,B\} $, where $\sum_{k=1}^\delta n_k=N_{\emptyset}$, the weight of the configuration can be written as
\begin{equation}
\label{Weights ZRP}
W_{ZRP}\bigl(\{n_{1}\tau_{1},\cdots,n_{\delta}\tau_{\delta}\}\bigr)
=\prod_{k=1}^{\delta} f_{\tau_{k}}(n_{k})\,,
\end{equation}
where $f_{A}(n)$ ($f_{B}(n)$) is the weight of an $A(B)$-box  containing $n$ particles. In order to compute $f_{A}(n)$ and $f
_{B}(n)$, let us define the vectors $\vert a_{1} \rangle$, $ \langle a
_{2} \vert$, $\vert b_{1} \rangle$ and $\langle b_{2} \vert$  by
\begin{equation}
\label{VECTOR}
 \vert a_{1} \rangle= \vert b_{1} \rangle= \vert 1 \rangle+ \vert 2 \rangle,\quad  \langle a_{2} \vert=\langle 1\vert,\quad \langle b_{2} \vert
=p\langle 2\vert\,,
\end{equation}
where we used the basis vectors
 \begin{equation}
\label{BaseVectors}
\vert 1 \rangle=\left(\begin{array}{c}
  1\\ 0
  \end{array}\right) \; , \;
\vert 2\rangle=\left(\begin{array}{c}
  0 \\ 1
  \end{array}\right)\,.
\end{equation}
Then the operators $\mathbf A$ and $\mathbf B$ in the matrix representation~(\ref{MATRIX A,B,E}) can be rewritten as 
\begin{equation}
\label{REWITEA,B}
\mathbf A=  \vert a_{1} \rangle\langle a_{2} \vert,\qquad
\mathbf B
=  \vert b_{1} \rangle\langle b_{2} \vert.
\end{equation}
Using Eqs.~(\ref{VECTOR})-(\ref{REWITEA,B}) and~(\ref{MATRIX A,B,E}) we obtain
\begin{align}
\label{WeightA}
f_{A}(n)&=\langle a_{2} \vert \mathbf E^{n}\vert a_{1} \rangle=\langle a_{2} \vert \mathbf E^{n}\vert b_{1} \rangle=1\\
\label{WeightB}
f_{B}(n)&=\langle b_{2} \vert \mathbf E^{n}\vert b_{1} \rangle=\langle b_{2} \vert \mathbf E^{n}\vert a_{1} \rangle=p\alpha^{n}.
\end{align}
We can show that Eq.~(\ref{Weights ZRP}) satisfy the pairwise balance condition \cite{Pairwise Balance}, therefore it is the stationary state for the dynamics specified by~(\ref{RulesZRP})-(\ref{EndRulesZRP}). 

Let us finally turn to the case $p=0$. It is clear from Eqs.~(\ref{Weights ZRP}),~(\ref{WeightA}) and~(\ref{WeightB}) that the stationary state weight of the ZRP consists only of the weights of $A$-boxes containing particles. Defining $\langle N
_{A}\rangle$ as the average number of $A$-boxes and $\langle n\rangle$ as the average number of particles in an $A$-box, and noticing the dynamical rules of the non-conserving ZRP,  (\ref{A particlesDensityP=0}) and (\ref{VacanciesDensityP=0}), we observe different behaviors for $\langle N_{A}\rangle$ and $\langle n\rangle$, namely
\begin{itemize}
\item
 for $p=0,\, \alpha<\alpha_{c}$, $\langle N_{A}\rangle$  and $\langle n\rangle$ are finite.
\item
 for $p=0,\, \alpha>\alpha_{c}$, $\langle N_{A}\rangle={\mathcal O}(1)$ and $\langle n\rangle={\mathcal O}(L)$. 
\end{itemize}
Therefore, we have a condensation transition where a large number of particles accumulate in a single $A$-box.
 \color{black}


\section{Numerical results}

\label{sec:numerics}
Since all stationary properties of the model defined in (\ref{DYNAMICALRULES}) can be computed exactly, our numerical simulations focus on its dynamical evolution. As we will see, the dynamical behavior is affected by strong scaling corrections which will be explained heuristically in Sect.~\ref{sec:toymodel}.

\begin{figure}
\centering\includegraphics[width=\linewidth]{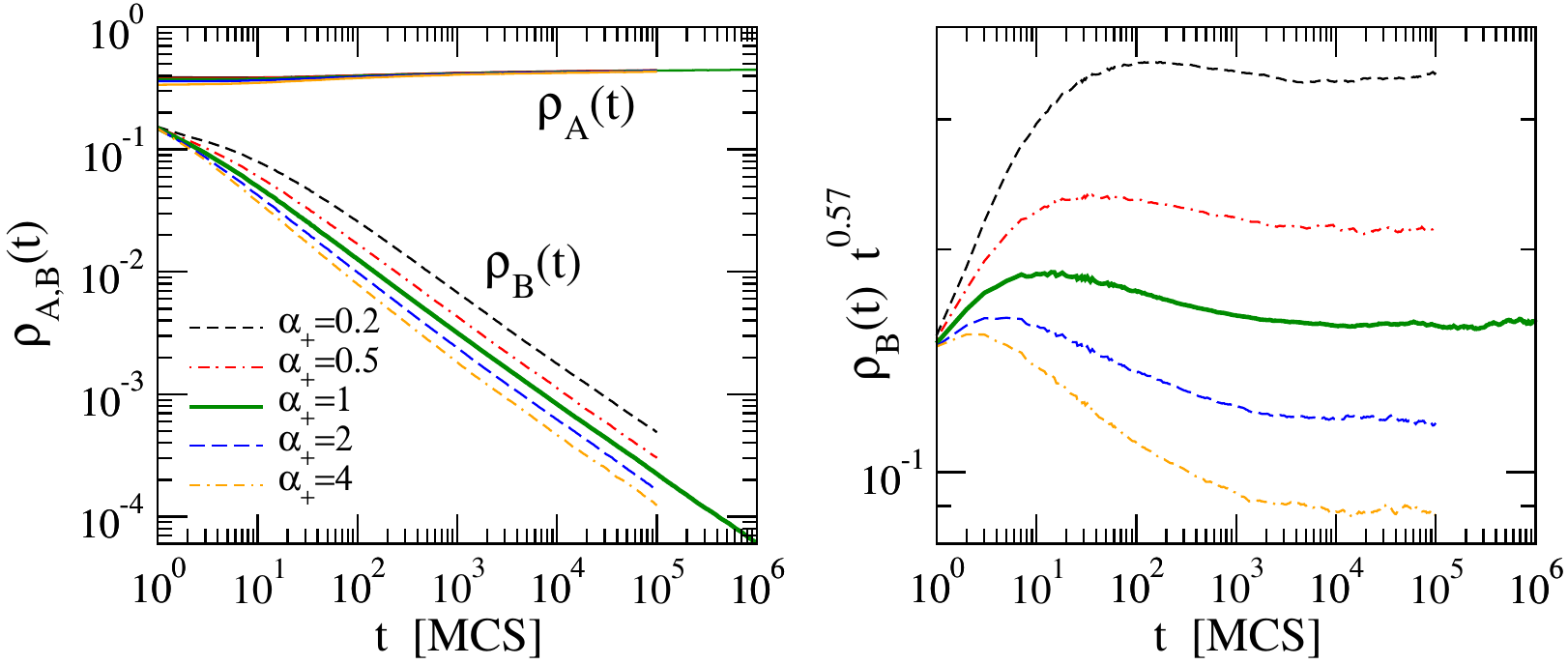}
\vspace*{-6mm}
\caption{Decay of the order parameters $\rho_{A,B}$ at the critical point in a very large system with $10^5$ sites with random initial conditions (see text).}
\label{fig:num-decay}
\end{figure}

\subsection{Decay of $\rho_A$ and $\rho_B$ at the critical point}
%
At the critical $p=0$, $\alpha=2$ we have $p_+/p_-=p=0$ and $\alpha_+/\alpha_-=2$, implying $p_+=0$, meaning that at this point the model controlled by two parameters $\alpha_+$ and $p_-$. In Fig.~\ref{fig:num-decay} we measured the time dependence of both order parameters for $p_-=1$ and various values of $\alpha_+$, starting with a random initial state with $\rho_A(0)=\rho_B(0)=1/3$. The behavior turns out to be  qualitatively similar in all cases: While the density $\rho
_A(t)$ seems to increase slightly, the density $\rho_B(t)$ shows a decay reminding of a power law $\rho_B(t)\sim t^{-\delta}$. However, if we first estimate the exponent $\delta\approx 0.57$ and then divide the data by $t^{-\delta}$ one observes a significant curvature of the data: The effective exponent $\delta_{\rm eff}$ decreases from 0.6 down 0.57 without having reached a stable value in the numerically accessible regime, indicating strong scaling corrections.

It turns out that the effective exponent depends strongly on the particle densities in the initial state. This freedom can be used to reduce the influence of the scaling corrections. Choosing for example a random initial configuration with $\rho_A(0)=0.9$ and $\rho_B(0)=0.1$ one obtains a less pronounced curvature of $\rho_B(t)$ with an effective exponent of only $\delta\approx 0.51$. This suggests that the asymptotic exponent is $\delta
=1/2$.

\subsection{Finite-size scaling}
\begin{figure}
\centering\includegraphics[width=\linewidth]{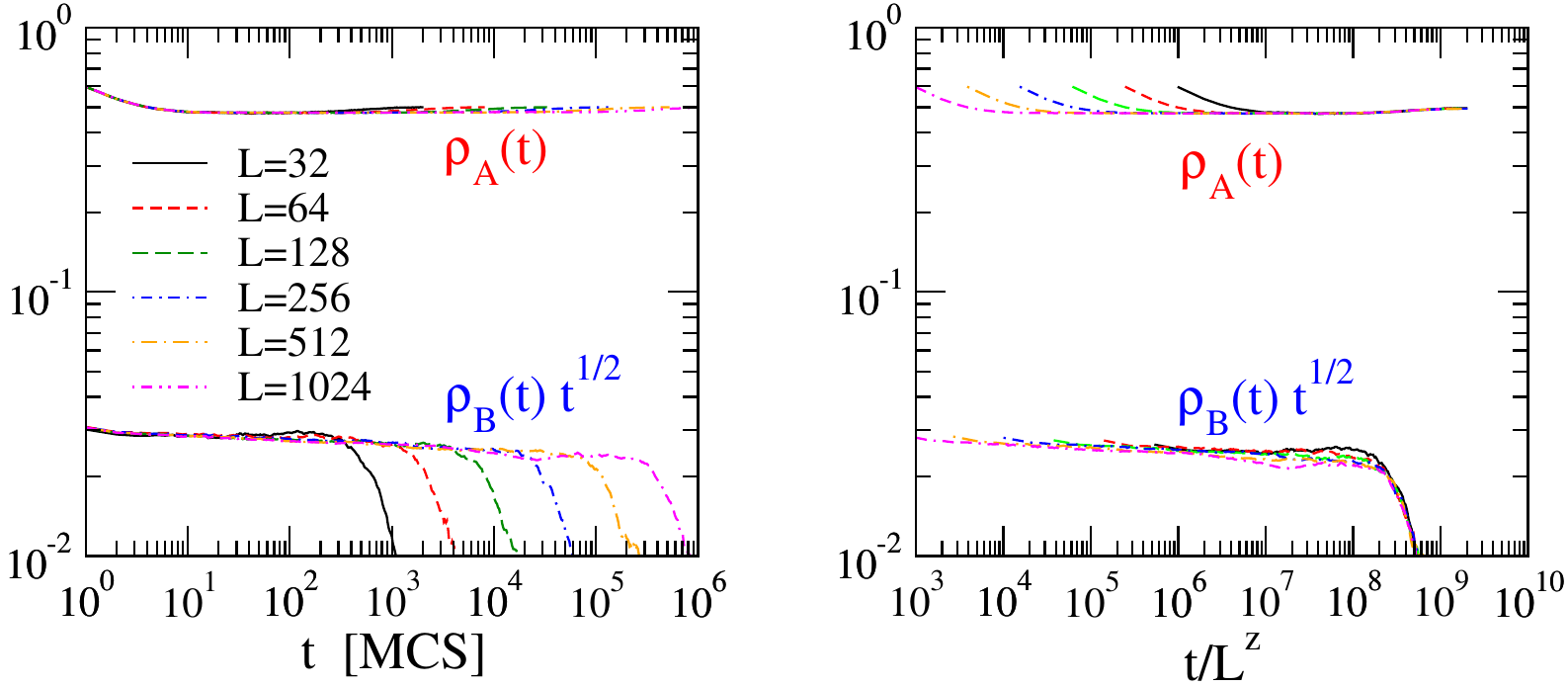}
\vspace*{-6mm}
\caption{Finite-size scaling at the critical point (see text).}

\label{fig:num-fs}
\end{figure}

%
Using the initial condition $\rho_A(0)=0.9$ and $\rho_B(0)=0.1$ we repeated the simulation in finite systems. The results are plotted in the left panel of Fig.~\ref{fig:num-fs}, where we divided $\rho_B(t)$ by the expected power law $t^{-1/2}$ so that an infinite system should produce an asymptotically horizontal line. As can be seen, a finite system size leads to a sudden breakdown of $\rho_B(t)$ while there is no change in $\rho_A(t)$. Plotting the same data against $t/L^z$ (right panel), where $z
=\nu_\parallel/\nu_\perp$ is the dynamical exponent, the best data collapse is obtained for $z=2$. This is plausible since so far all systems, which have been solved by means of matrix product methods, are essentially diffusive with a dynamical exponent $z=2$.

\subsection{Off-critical simulations}

\begin{figure}
\centering\includegraphics[width=\linewidth]{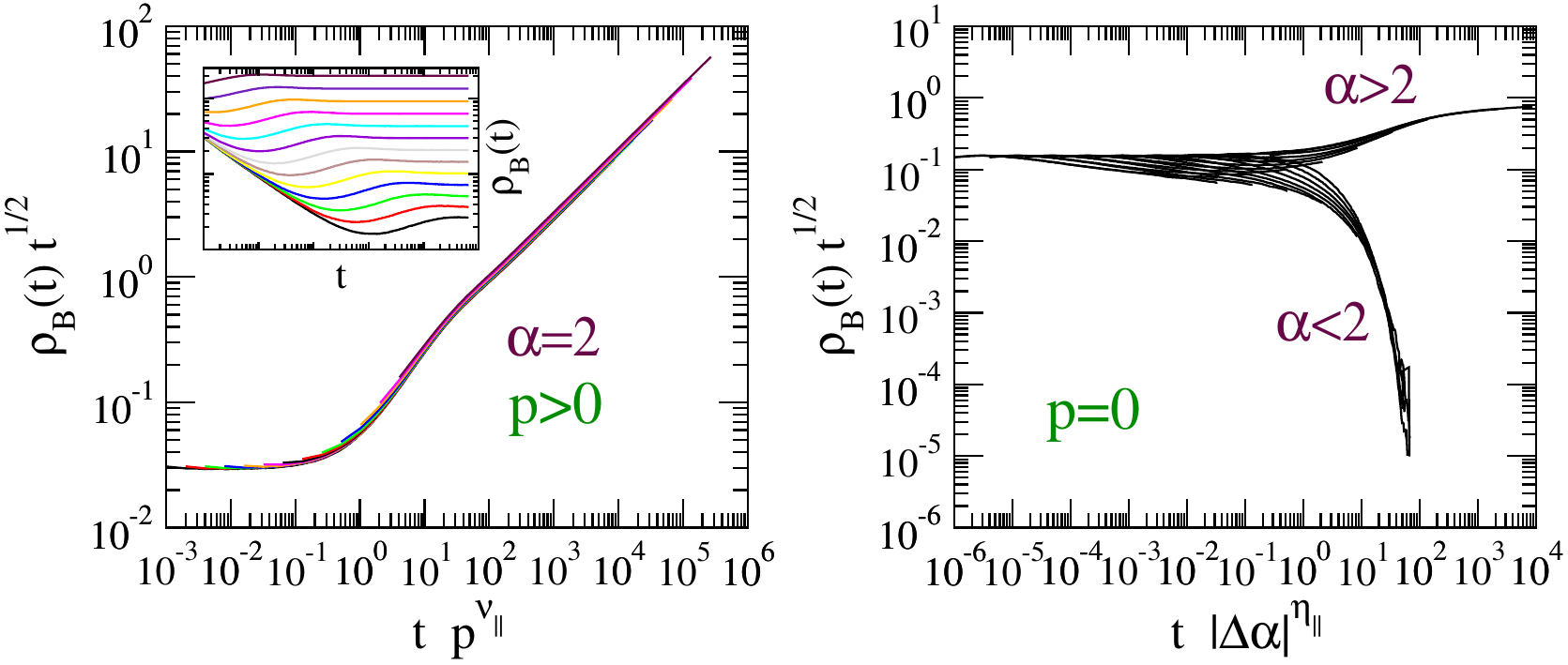}
\vspace*{-6mm}
\caption{Data collapses for off-critical simulations. Left: Variation of $p$ in the range $0.0001,0.0002,\ldots,0.4096$. The inset shows the corresponding raw data. Right: Variation of $\Delta \alpha=\alpha-2$ in the range $\pm 0.001$, $\pm 0.002$, $\ldots \pm 0.512$. }
\label{fig:num-offcrit}
\end{figure}

Finally we investigate the two-dimensional vicinity of the critical point where 
\begin{equation}
\Delta\alpha \;=\; \alpha-\alpha_c \;=\; \alpha-2
\end{equation}
as well as $p$ are small. First we choose $\Delta\alpha=0$ and study the model for $p>0$. In this case the order parameter $\rho_B(t)$ first decays as if the system was critical until it saturates at a constant value, as shown in the inset of Fig.~\ref{fig:num-offcrit}. 

Surprisingly, $\rho_B(t)$ first goes through a local minimum and then increases again before it reaches the plateau. This phenomenon of \textit{undershooting} has also been observed in conserved sandpile models~\cite{SandpilesDP} and may indicate that the system has a long-time memory for specific correlations in the initial state. Plotting $\rho_B(t) t^{1/2}$ against $t p^{\nu_\parallel}$ one finds an excellent data collapse for $\nu_\parallel=1.00(5)$, indicating that $\nu_\parallel=1$.

Next we keep $p=0$ fixed and vary $\Delta\alpha$. For $\Delta\alpha<0$ one finds that the density $\rho_B(t)$ crosses over to an exponential decay. For $\Delta\alpha>0$, where one expects supercritical behavior, $\rho_B(t)$ does \textit{not} saturate at a constant, instead it first decreases as $t^{-1/2}$ followed by a short period of a decelerated decay until it continues to decay as $t^{-1/2}$. This means that $\alpha>2$ causes an increase of the amplitude but not a crossover to a different type of decay. To our knowledge this is the first example of a power law to the same power law but with a different amplitude.

Plotting $\rho_B(t) t^{1/2}$ against $t p^{\eta_\parallel}$ the data collapse is unsatisfactory due to the scaling corrections discussed above. However, the best compromise is obtained for $\eta_\parallel=1.9(2)$, which is compatible with $\eta_\parallel=2$.

\subsection{Phenomenological scaling properties}
%
Apart from the scaling corrections which will be discussed in the following section, the collected numerical results suggest that the process in the vicinity of the critical point is invariant under scale transformations of the form
\begin{align}
\label{ScalingScheme}
&t \to \Lambda^{\nu_\parallel} t\,, \qquad L \to \Lambda^{\nu_\perp} L\,, \qquad \rho_B \to \Lambda^{\beta}\rho_B  \notag\\
&p \to \Lambda p \,, \qquad \Delta \alpha \to \Lambda^\theta \Delta\alpha\,,
\end{align}
where $\theta=\nu_\parallel/\eta_\parallel$ is the crossover exponent between the two control parameters. 

Assuming that the critical behavior is described by simple rational exponents, our findings suggest that the universality class of the process is characterized by four exponents $\beta=1/2\,,\quad \nu_\parallel=1\,,\quad \nu_\perp=1/2 \,,\quad \theta=1/2$ together with the scaling relations
\begin{align}
&\delta = \frac{\beta}{\nu_\parallel}=\frac12 \\
&z
= \frac{\nu_\parallel}{\nu_\perp} = \frac{\eta_\parallel}{\eta
_\perp}=2\\
&\theta=\frac{\nu_\parallel}{\eta_\parallel}=1/2\,.
\end{align}
The values of the exponents are listed in Table~\ref{tab:exponents}. Regarding the stationary properties for $p>0$, these exponents are in full agreement with the exact solution in Sect.~\ref{sec:exact}.

The scaling scheme (\ref{ScalingScheme}) implies various scaling relations. For example, it allows us to predict that the stationary density of $B$-particles in the vicinity of the critical point should scale as
\begin{equation}
\label{StationaryScaling}
\rho_B^{\rm stat} \;=\; p^\beta F\Bigl(\frac{(\Delta\alpha)^2}{p}\Bigr)\,,
\end{equation}
where $F$ is a universal scaling function. Comparing this form with the exact result (\ref{B particlesDensityTrmo}) we find that 
\begin{equation}
F(\xi)=\frac{1}{2 \sqrt{4+\xi}}\,.
\end{equation}

\begin{table}
\begin{center}
\begin{tabular}{|c|c|c|c|c|c|c|c|}
\hline
$\quad\beta\quad$ & $\quad\nu_\perp\quad$ & $\quad\nu_\parallel\quad$ & $\quad z\quad$ & $\quad\eta_\perp\quad$ & $\quad\eta_\parallel\quad$ & $\quad\theta\quad$ & $\quad\delta\quad$\\
\hline
$1/2$ & $1/2$ & $1$ & $2$ & $1$ & $2$ & $1/2$ & $1/2$\\
\hline
\end{tabular}
\end{center}
\vspace{-3mm}
\caption{Expected values of the critical exponents. \label{tab:exponents}}
\end{table}


\section{Heuristic explanation of the critical behavior} 

\label{sec:toymodel}

\subsection{Reduction to an effective model}
%
The model investigated above can be related to an effective process of pair-creating and annihilating random walks. As we will see below, this effective model captures the phase structure and the essential critical properties of the full model. 

\begin{figure*}[t]

\centering\includegraphics[width=0.9\linewidth]{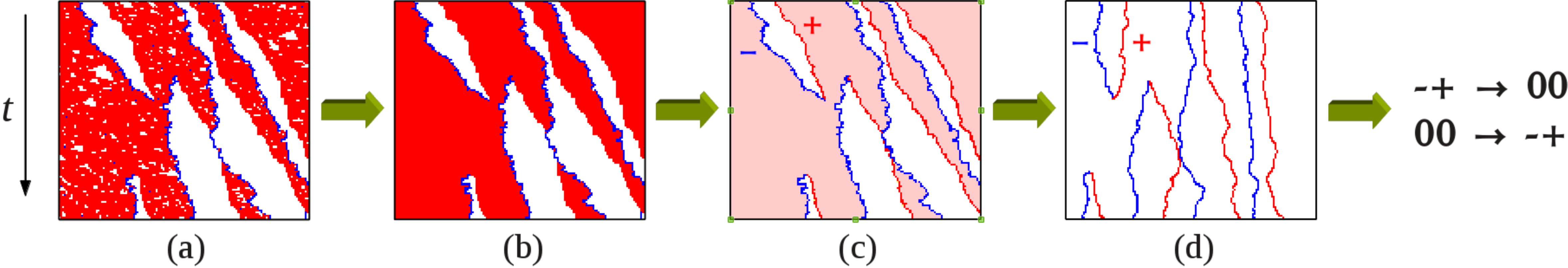}
\vspace*{-3mm}
\caption{Motivation of the reduced model (see text). (a) Temporal evolution of the original model for $\alpha=2$ and $p=0.01$ in a vertically compressed representation with 20 Monte-Carlo sweeps per pixel. As before, particles of type $A$ and $B$ are marked by red and blue pixels, respectively. (b) Illustration of compactified $A$-domains. (c)~Kink representation, interpreted as a pair-creating and annihilating diffusion process. (d) Removal of the overall bias. }
\label{fig:toymodel}
\end{figure*}

Starting point is the observation that the original model, especially close to the critical point, tends to form dense and sharply bounded domains of $A$-particles while the $B$-particles are sparsely distributed. The $A$-domains are not compact, rather they are interspersed by little patches of empty sites. As can be seen in Fig.~\ref{fig:toymodel}a, these small voids inside the $A$-domains do not exceed a certain typical size. This suggests that they can be regarded as some kind of local noise which is irrelevant for the critical behavior on large scales, meaning that we may disregard them and consider the $A$-domains effectively as compact objects, as shown schematically in Fig.~\ref{fig:toymodel}b. 

Secondly we note that the $B$-particles in the full model are predominantly located at the right boundary of the $A$-domains. This suggests that the dynamics can be encoded effectively in terms of the left and right boundaries of the $A$ domains, interpreted as charges $-$ and~$+$ (see Fig.~\ref{fig:toymodel}c). In this kink representation, the negative charges can be identified with the $B$-particles in the original model, while the positive charges can be understood as marking the left boundary of $A$-domains.

Thirdly, we observe that the dynamics of the original model is biased to the right. In the kink representation, an overall bias does not change the critical properties of the model and can be eliminated in a co-moving frame, as sketched schematically in Fig.~\ref{fig:toymodel}d.

Having completed this sequence of simplifications, the original process can be interpreted as an effective pair-creating and annihilating random walk of $+$ and $-$ charges according to the reaction-diffusion scheme
\begin{align}
\label{toyrules}
+\emptyset \stackrel{\lambda}\longrightarrow \emptyset + \qquad \emptyset + \stackrel{1/\lambda}\longrightarrow +\emptyset\notag\\
-\emptyset \stackrel{1/\lambda}\longrightarrow \emptyset - \qquad \emptyset - \stackrel{\lambda}\longrightarrow -\emptyset\\
-+ \stackrel{1}\longrightarrow \emptyset\emptyset\qquad\;\;
\emptyset\emptyset \stackrel{q}\longrightarrow -+\notag
\end{align}
Here the parameter $\lambda$ controls the relative bias between the two particle species and thus it is expected to play the same role as $\alpha$ in the full model, although with a different critical value $\lambda_c=1$. The other parameter $q$ controls the rate of spontaneous pair creation and therefore plays a similar role as $p$ in the original model.

The reduced process starts with an alternating initial configuration $+-+-+-...$, where $\rho_+(0)=\rho_-(0)=1/2$. As time evolves, particles are created and annihilated in pairs, meaning that the two densities 
\begin{equation}
\rho_+(t)
= \rho_-(t)\,
\end{equation}
are exactly equal. These densities are expected to play the same role as the order parameter $\rho_B(t)$ in the original model.

\subsection{Numerical results for the reduced model}
%
The reduced model has the advantage that it can be implemented very efficiently on a computer by storing the coordinates of the kinks in a dynamically generated list. Simulating the model we find the following results:
\begin{itemize}
\item
$q>0$: The model evolves into a stationary state with a constant density $\rho_{+}=\rho_-$, qualitatively reproducing the corresponding results for the full model shown in the right panel of Fig.~\ref{fig:rho}. 
\item 
$q=0, \,\lambda>1$: Positive charges move to the right and negative charges move to the left until they form bound $+-$ pairs which perform a slow unbiased random walk. If two such pairs collide they coagulate into a single one by the effective reaction $+-+-\to +-$. Therefore, one expects the density of particles to decay as $t^{-1/2}$ in the same way as in a coagulation-diffusion process~\cite{Coagulation}.
\item
$q=0,\, \lambda=1$: At the critical point the particle density seems to decay somewhat faster than $t^{-1/2}$. The origin of these scaling corrections will be discussed below.
\item 
$q=0, \,\lambda<1$: In this case the negative charges diffuse to the right while positive charges diffuse to the left. When they meet they quickly annihilate in pairs, reaching an empty absorbing state in exponentially short time.
\end{itemize}
%
\begin{figure}
\centering\includegraphics[width=\linewidth]{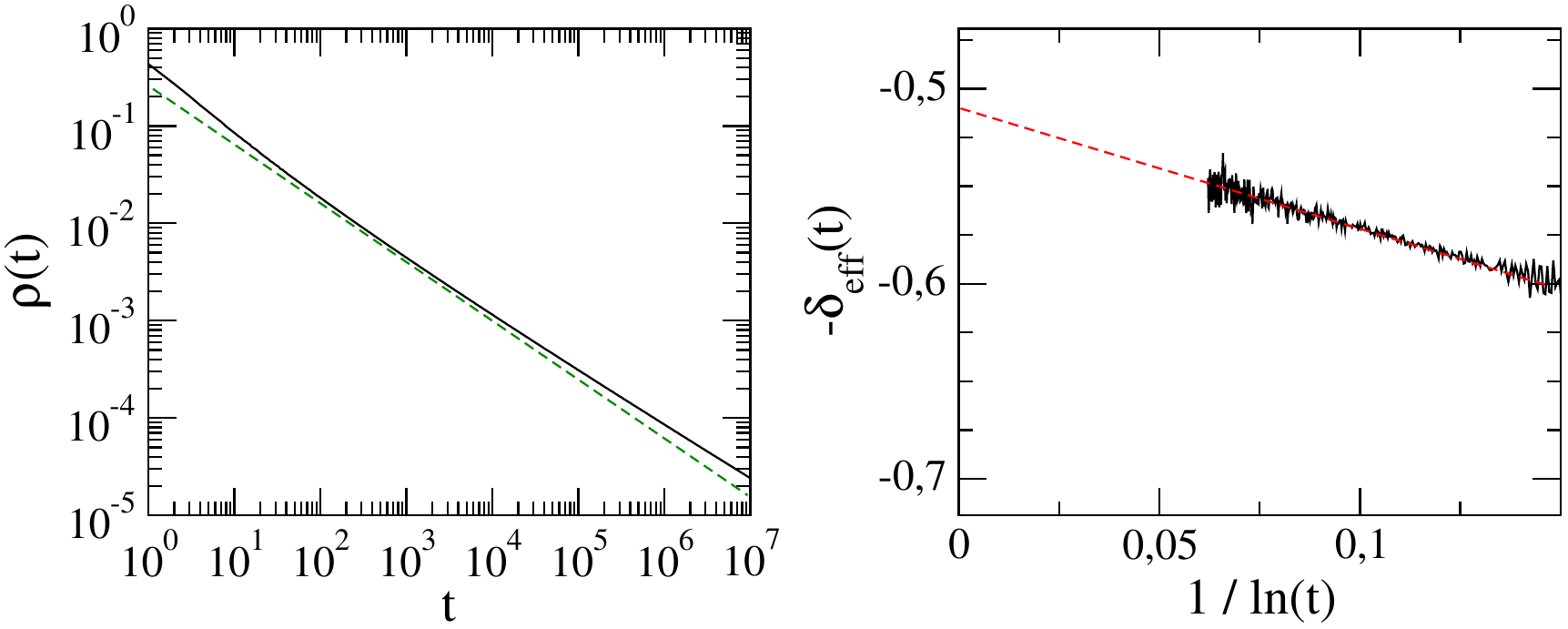}
\vspace{-7mm}
\caption{Numerical simulation of the reduced model with $L=10^7$ sites simulated at the critical point. Left panel: Decay of the particle density $\rho(t)$. The green dashed straight line visualizes the slow curvature of the data, indicating persistent scaling corrections. Right panel: Corresponding local slopes plotted against $1/\ln(t)$, interpreted as an effective critical exponent $-\delta_{\rm eff}(t)$. A visual extrapolation along the red dashes line to $t\to \infty$ is consistent with the expected asymptotic exponent $\delta=0.5$.}
\label{fig:decay}
\end{figure}
%
Therefore, the reduced model exhibits the same type of critical behavior as the full model. Moreover, repeating the standard simulations of Sect.~\ref{sec:numerics} (not shown here) we obtain similar estimates of the critical exponents.

\subsection{Explaining the scaling corrections heuristically}
%
Performing extensive numerical simulations of the reduced model at the critical point over seven decades in time (see Fig.~\ref{fig:decay}) one can see a clear curvature in the double-logarithmic plot. Unlike initial transients in other model, this curvature seems to persist over the whole temporal range. To confirm this observation, we plotted the corresponding local exponent $\delta_{\rm eff}$ against $1/\ln(t)$ in the right panel of the figure. If the curve is extrapolated visually to $t \to \infty$, the most likely extrapolation limit is indeed $\delta=1/2$, confirming our previous conjecture in the case of the full model.

Where do the slow scaling corrections come from? This question is of general interest because various other nonequilibrium phase transitions, where the universal properties are not yet fully understood, show similar corrections. For example, the diffusive pair contact process~\cite{PCPD} and fixed-energy sandpiles~\cite{MannaDP,SandpilesDP} both exhibit a similar slow curvature of the particle decay at the critical point. Here we have a particularly simple system with an exactly known critical point, where the origin of the slow scaling corrections can be identified much easier.

To explain the scaling corrections heuristically, let us consider the pair annihilation process defined in (\ref{toyrules}) at the critical point starting with an alternating initial configuration ($+-+-+-...$). We first note that this process has the special property that pairs of particles which eventually annihilate must have been nearest neighbors in the initial configuration. In so far this process differs significantly from the usual annihilation process $2A\to \emptyset$, where in principle any pair can annihilate.

If the process had started with only a single $-+$ pair, both particles would perform a simple random walk until they collide and annihilate. In this case the annihilation probability would be related to the first-return probability of a random walk~\cite{SidRedner}. Since the first-return probability is known to scale as $t^{-3/2}$ in one spatial dimension, the life time of the pair, which is obtained by integration over time, would decay as $t^{-1/2}$. However, in the present case the $-+$ pair is interacting with other pairs to the left and to the right. These neighboring pairs impose a kind of non-reactive fluctuating boundary, limiting the space in which the random walk of the two particles can expand. In other words, the neighboring pairs lead to a small effective force, pushing the two charges towards each other. This in turn enhances the frequency of annihilation events, explaining qualitatively why the particle density first decays faster than $t^{-1/2}$.

\begin{figure}
\centering\includegraphics[width=0.8\linewidth]{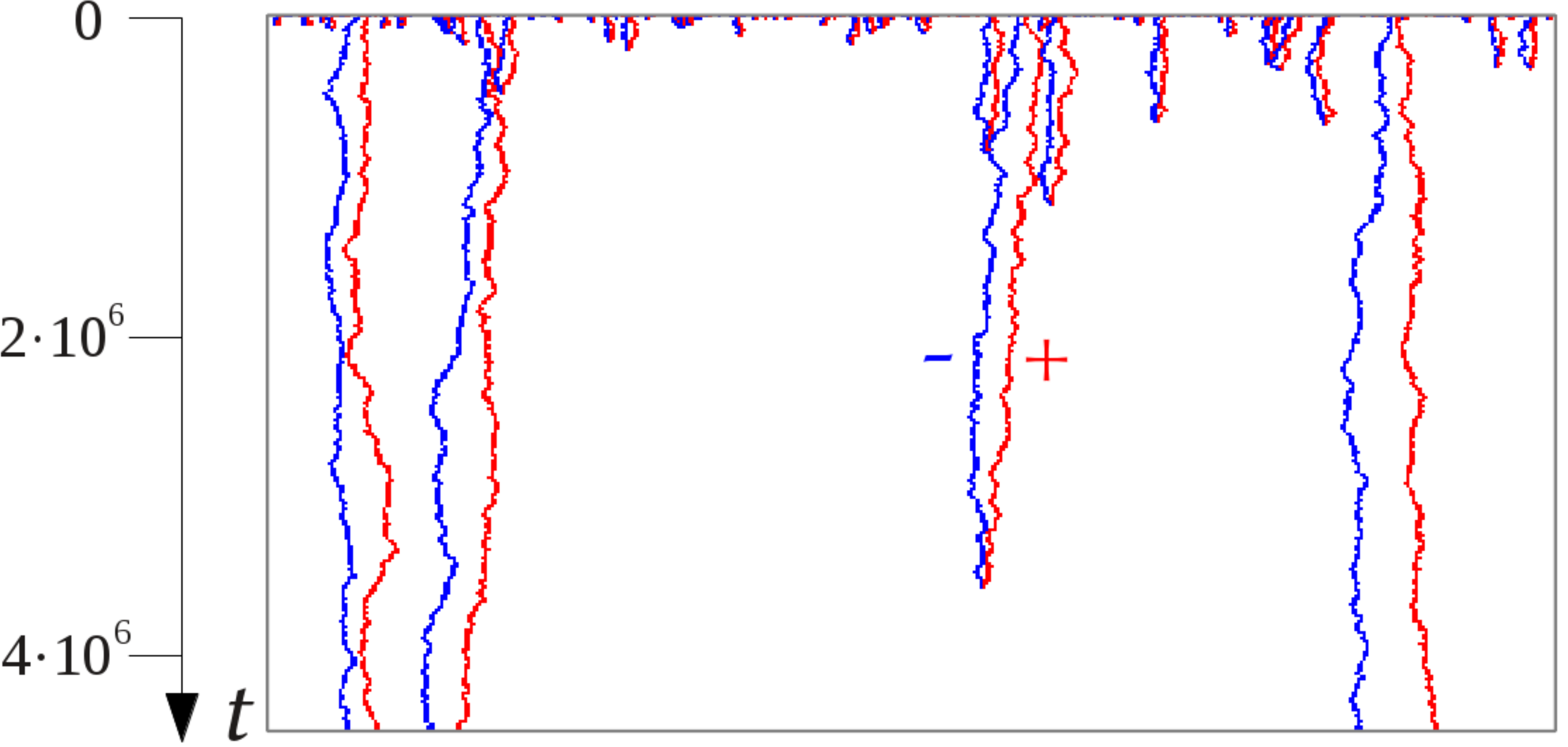}
\centering\includegraphics[width=\linewidth]{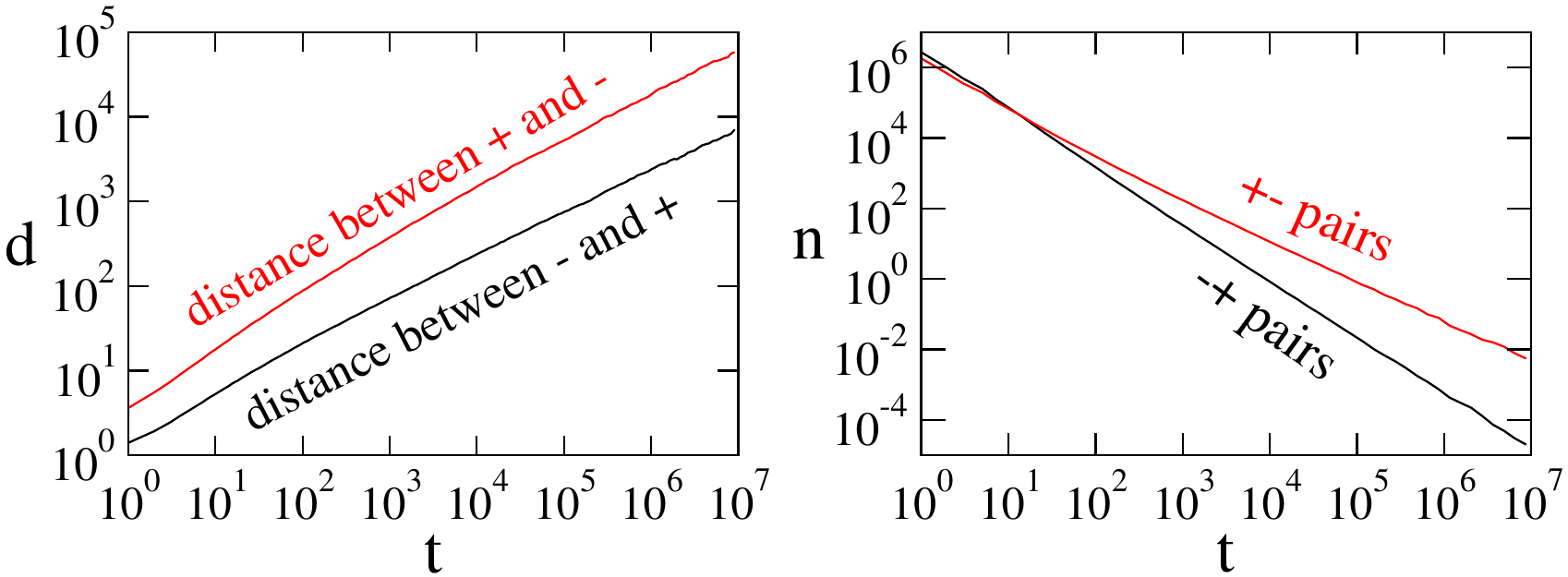}
\vspace{-5mm}
\caption{Explanation of the accelerated decay in the reduced charge model. The upper panel shows a typical snapshot of the process at the critical point monitored over long time. As can be seen, the process preferentially forms $-+$ pairs separated by large empty intervals. This impression is confirmed by a measurement of the average distance between neighboring charges shown in the lower left panel. Likewise, the average number of adjacent $+-$ and $-+$ pairs evolves differently.}
\label{fig:expl}
\end{figure}

However, as time proceeds the accelerated decay of the particle density leads to a corresponding increase of the average distance between the particles which grows faster than $t^{1/2}$. Since the average distance between $-$ and $+$ particles cannot grow faster than $t^{1/2}$, this implies that the average distance between $+$ and $-$ has to grow faster than $t^{1/2}$, as we could confirm by numerical measurements in Fig.~\ref{fig:expl}. This in turn implies that the effective force mentioned above decreases with time.

%

To find out how fast the effective force decreases with time, we first note that the force is caused by adjacent $+-$ pairs which cannot penetrate each other. A numerical measurement shows that the number of $+-$ pairs decays in the same way as the squared particle density, i.e. like in a mean-field approximation (see right panel of Fig.~\ref{fig:expl}), while the number of $-+$ pairs is -- as expected -- proportional to the particle loss:
\begin{equation}
n_{+-}(t) \sim \rho^2(t) \,, \qquad n_{-+}(t) \sim \dot\rho(t).
\end{equation}
Therefore, we expect the effective force to be proportional to $\rho^2(t)$ which roughly scales as $t^{-1}$. Thus we conclude that the particle density of the pair annihilation process at the critical point  (and similarly in the full model) decays in the same way as the survival probability of a one-dimensional random walk starting at the origin subjected to a time-dependent bias proportional to $1/t$ towards the origin, terminating upon the first passage of the origin. In fact, simulating such a random walk we find slowly-decaying logarithmic corrections of the same type, confirming the heuristic arguments given above. To our knowledge an exact solution of a first-passage random walk with time-dependent bias is not yet known.
\section{Conclusions}

\label{sec:conclusions}

In this work we have introduced and studied a two-species reaction-diffusion process on a one-dimensional periodic lattice which exhibits a nonequilibrium phase transition. Its stationary state can be determined exactly by means of the matrix product method. Together with numerical studies of the dynamics we have identified the critical exponents which are listed in Table~\ref{tab:exponents}. The transition can be explained qualitatively by relating the model to a reduced process (see Sect.~\ref{sec:toymodel}). This relation also provides a heuristic explanation of the unusual corrections to scaling observed in this model.

Our findings seem to be in contradiction with a previous claim by one of the authors~\cite{Comment,Book} that first-order phase transitions in non-conserving systems with fluctuating domains should be impossible in one dimension. In \cite{Comment} it was argued that a first-order transition needs a robust mechanism in order to eliminate spontaneously generated minority islands of the opposite phase, but this would be impossible in 1D because in this case the minority islands do not have surface tension. Although this claim was originally restricted to two-state models, the question arises why we find the contrary in the present case.

Again the caricature of the reduced process sketched in Fig.~\ref{fig:toymodel}a provides a possible explanation: As can be seen there are two types of white patches, namely, large islands with a blue $B$-particle at the left boundary, and small islands without. This means that the $B$-particles are used for marking two different types of vacant islands, giving them different dynamical properties. Only the large islands containing a $B$-particle are minority islands in the sense discussed in~\cite{Comment}, while the small islands without $B$-particles inside the $A$-domains are biased to shrink by themselves. 

Therefore, we arrive at the conclusion that first-order phase transitions in non-conserving 1D systems with fluctuating domains are indeed possible in certain models with several particle species if one of the species is used for marking different types of minority islands.
\appendix
\section{An exactly solvable three species model}

\label{sec:appendix}

 In this appendix, we show that a similar type of phase transition can also exist
in four-state models. We introduce an exactly solvable one-dimensional driven-diffusive model with non-conserved dynamics consisting of three species of particles. The system evolves random-sequentially according to the dynamical rules~(\ref{DYNAMICAL RULE})
where $I, J,K\in \{A,B,C\} $. This system is defined by the processes
\begin{equation}
\label{DYNAMICAl3PARTICLES}
\begin{array}{ccc}
A\emptyset\mathop{\rightarrow}\limits^{\lambda_{+}} \emptyset A,&B\emptyset\mathop{\rightarrow}\limits^{\alpha_{+}} \emptyset B,&
C\emptyset\mathop{\rightarrow}\limits^{\beta_{+}} \emptyset C\\
AA\mathop{\rightleftharpoons}\limits^{p_{+}}_{p_{-}} BA,&AC\mathop{\rightleftharpoons}\limits^{p_{+}}_{p_{-}} BC,&
AB\mathop{\rightleftharpoons}\limits^{p_{+}}_{p_{-}} BB\\
AB\mathop{\rightleftharpoons}\limits^{q_{+}}_{q_{-}} CB,&AA\mathop{\rightleftharpoons}\limits^{q_{+}}_{q_{-}} CA,&
AC\mathop{\rightleftharpoons}\limits^{q_{+}}_{q_{-}} CC\\
BA\mathop{\rightleftharpoons}\limits^{q}_{p} CA,&BB\mathop{\rightleftharpoons}\limits^{q}_{p} CB,&
BC\mathop{\rightleftharpoons}\limits^{q}_{p} CC\\
AB\mathop{\rightleftharpoons}\limits^{\lambda_{+}}_{\alpha_{+}} B\emptyset,&AC\mathop{\rightleftharpoons}\limits^{\lambda_{+}}_{\beta_{+}} C\emptyset,&
CB\mathop{\rightleftharpoons}\limits^{\alpha}_{q} B\emptyset\\
BB\mathop{\rightleftharpoons}\limits^{\alpha}_{p} B\emptyset,&BC\mathop{\rightleftharpoons}\limits^{\beta}_{p} C\emptyset,&
CC\mathop{\rightleftharpoons}\limits^{\beta}_{q} C\emptyset\\
A\emptyset\mathop{\rightleftharpoons}\limits^{p_{+}}_{p_{-}} BA,&A\emptyset\mathop{\rightleftharpoons}\limits^{1}_{1} AA,&
A\emptyset\mathop{\rightleftharpoons}\limits^{q_{+}}_{q_{-}} CA\\
\end{array}
\end{equation}
where the rates $\alpha$, $\beta$, $p$ and $q$ are given by the ratios
$$\alpha=\frac{\lambda_+}{\alpha_+}\,,\qquad \beta=\frac{\lambda_+}{\beta_+}\,,\qquad
p = \frac{p_+}{p_-}\,,\qquad q = \frac{q_+}{q_-}.$$
The first four lines of~(\ref{DYNAMICAl3PARTICLES}) have been studied in Ref.~\cite{Basu} where the phase transition is not accessible. We have found that the  matrix algebra of the dynamical rules~(\ref{DYNAMICAl3PARTICLES}) has a three-dimensional matrix representation given by the following matrices 
\begin{equation}
\label{MATRIX A,B,C,E}
\begin{array}{l}
\mathbf A=\left(
\begin{array}{ccc}
1 & 0 & 0\\
1 & 0 & 0\\
1 & 0 & 0
\end{array} \right),\;
\mathbf B=p\left(
\begin{array}{ccc}
0 & 1 & 0\\
0 & 1 & 0\\
0 & 1 & 0
\end{array} \right), \\\\
\mathbf C=q\left(
\begin{array}{ccc}
0 & 0 & 1\\
0 & 0 & 1\\
0 & 0 & 1
\end{array} \right),\;
\mathbf E=\left(
\begin{array}{ccc}
1 & 0 & 0\\
0 & \alpha & 0\\
0 & 0 & \beta
\end{array} \right).
\end{array}
\end{equation}
The representation~(\ref{MATRIX A,B,C,E}) is the same as the matrix representation represented in Ref.~\cite{Basu}.  
The partition function defined as a sum of the weights of all available configurations with at least one particle, is given by
\begin{equation}
\label{PartitionFunction3Particles}
Z_{L}={\rm Tr}\bigl[(\mathbf A+\mathbf B+\mathbf C+\mathbf E)^L-\mathbf E^L\bigr]\,.
\end{equation}
The stationary density of the A, B and C-particles can be written as
\begin{equation}
\label{A particlesDensity3Particles}
\rho_A^{stat}=\frac{{\rm Tr}\bigl[\mathbf A(\mathbf A+\mathbf B+\mathbf C+\mathbf E)^{L-1}\bigr]}{Z_{L}},
\end{equation}
\begin{equation}
\label{B particlesDensity3Particles}
\rho_B^{stat}=\frac{{\rm Tr}\bigl[\mathbf B(\mathbf A+\mathbf B+\mathbf C+\mathbf E)^{L-1}\bigr]}{Z_{L}},
\end{equation}
\begin{equation}
\label{C particlesDensity3Particles}
\rho_C^{stat}=\frac{{\rm Tr}\bigl[\mathbf C(\mathbf A+\mathbf B+\mathbf C+\mathbf E)^{L-1}\bigr]}{Z_{L}}.
\end{equation}

We can compute the density of the vacancies using $\rho_{\emptyset}^{stat}=1-(\rho_A^{stat}+\rho_B^{stat}+\rho_C^{stat})$. Using the representation~(\ref{MATRIX A,B,C,E}) the equations~(\ref{PartitionFunction3Particles})-(\ref{C particlesDensity3Particles}) can be calculated exactly. In the thermodynamic limit $L\mathop{\rightarrow}\infty$, 
the density of the A-particles and  the vacancies vary discontinuously approaching the critical point, namely
\begin{enumerate}
\item[(i)]
 For $\beta\leq2$ and $p=q=0$, we find a discontinuous behavior as
$$\rho_A^{stat}=
\left\{
\begin{array}{ll}
\frac{1}{2}   
&  \mbox{for} \; \alpha <2\\ & \\
0
& \mbox{for} \;  \alpha >2,
\end{array}
\right.$$
$$\rho_{\emptyset}^{stat}=
\left\{
\begin{array}{ll}
\frac{1}{2}   
&  \mbox{for} \; \alpha <2\\ & \\
1
& \mbox{for} \;  \alpha >2,
\end{array}
\right.$$
and $\rho_B^{stat}=\rho_C^{stat}=0$.
\item[(ii)]
 For $\alpha\leq2$ and $p=q=0$, we find a discontinuous behavior as
$$\rho_A^{stat}=
\left\{
\begin{array}{ll}
\frac{1}{2}   
&  \mbox{for} \; \beta <2\\ & \\
0
& \mbox{for} \;  \beta >2,
\end{array}
\right.$$
$$\rho_{\emptyset}^{stat}=
\left\{
\begin{array}{ll}
\frac{1}{2}   
&  \mbox{for} \; \beta <2\\ & \\
1
& \mbox{for} \;  \beta >2,
\end{array}
\right.$$
and $\rho_B^{stat}=\rho_C^{stat}=0$.
\end{enumerate}


\end{document}